# Nonreciprocal Coupling Induced by Nonlocal Loss Engineering

*Yekai Shen[1,2][†], Shuhang Chen[2][†]\**

[1]School of the Gifted Young, University of Science and Technology of China, Hefei, China
[2]School of Information Science and Technology, University of Science and Technology of China, Hefei, China
\*Email address for corresponding author: chenshuhang@mail.ustc.edu.cn;
[†]These authors contributed equally of this work.

*Abstract*—Nonreciprocal coupling between photonic modes enables a range of advanced functionalities, though the available approaches for its practical implementation remain limited. Here, we introduce a novel strategy for achieving nonreciprocal coupling via nonlocal, nonlinear loss. We prove that robust, broadband, and continuously tunable nonreciprocal coupling can be realized by engineering the loss rate as a function of the state of a nonlocal mode, as validated through effective Hamiltonian modeling and numerical simulations. Our results suggest a promising route toward scalable, power-independent, and potentially integrable nonreciprocal photonic systems, with promising applications in non-Hermitian devices and topological photonics.

## I. Introduction

Nonreciprocity, a nontrivial asymmetric property of energy transport, has long served as a cornerstone of advanced photonics. It not only underpins key devices such as optical isolators, circulators, and switches [1], but has also recently emerged as a versatile tool in diverse areas, including topological photonics [2], neural networks [3], and optical computing [4]. Nonreciprocal coupling, wherein asymmetry is embedded directly in the interaction between two photonic modes, represents a more specialized form of nonreciprocity that offers unique perspectives into non-Hermitian physics [5-8]. In particular, such coupling is instrumental in realizing higher-order exceptional points, enabling the construction of Hatano–Nelson chains [5] that demonstrate the non-Hermitian skin effect [6] and enhanced nonreciprocal transport [7], as well as providing new opportunities for exploring unconventional topological physics [8].

Despite its importance, practical implementations for nonreciprocal coupling remains relatively limited. Experimental implementations to date have mainly relied on temporal modulation (including synthetic gauge fields) [7], nonlinear or optomechanical interactions [9, 10], and active gain schemes [11]; while these methods often involve extra clocking systems, strong input power, or amplifying elements. In this context, there is ongoing interest in new mechanisms for nonreciprocal coupling, offering advantages such as tunability, power independence, and potential integrability.

Recently, loss engineering has attracted increasing attention in wave physics [12-19]. By deliberately controlling dissipation, researchers have realized a variety of novel phenomena—such as loss-improved signal-to-noise ratios [12], loss-enhanced magneto-optical effects [13], loss-assisted metasurfaces [14], loss-enabled chirality inversion [15], and loss-modulated non-Hermitian topology [16]—transforming loss from an undesirable and unavoidable nuisance into a valuable degree of freedom for system control. Notably, the theoretical proposal of loss-induced nonreciprocity [17] provides a further approach for realizing nonreciprocal coupling, yet this strategy inherently assumes a priori asymmetry in the off-diagonal elements of the system Hamiltonian—a scenario rarely encountered in conventional physical systems. Nevertheless, this direction has inspired extensive theoretical and experimental efforts to implement nonreciprocity by loss engineering, including demonstrations in nonlinear-dissipative microresonators [18], hot atomic systems [19], and phononic crystals [20], while their application to photonic nonreciprocal coupling systems with large-scale integration and power-independent performance remains to be further explored.

In this work, we explore a mechanism for achieving nonreciprocal coupling via nonlocal loss engineering. We theoretically demonstrate that introducing nonlocal (i.e., dependent on the state of another mode) nonlinear loss to one of two coupled optical modes can enable the system to realize giant, tunable, power-robust, and broadband nonreciprocal coupling. We derive the effective Hamiltonian for this process and show through numerical simulations that the transmission characteristics of this system are equivalent to those of an ideal nonreciprocal coupling model. This mechanism leverages system loss in a nontrivial way, without relying on intrinsic asymmetry of the coupling coefficient, connecting modes, or multi-path coupling coherence, offering the potential for long-range wireless nonreciprocal coupling and large-scale integration in on-chip systems. Moreover, the degree of nonreciprocity can, in principle, be continuously tuned from reciprocal to fully nonreciprocal operation. These advantages may stimulate further research into novel physical phenomena and advanced applications, such as topological complex-energy braiding and high-capacity neural network emulation.

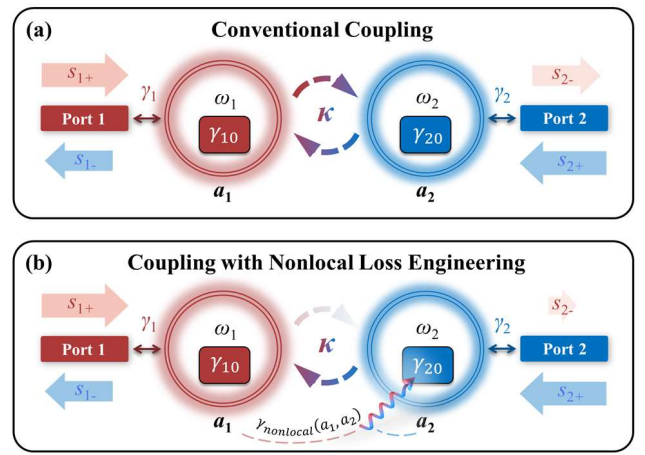

Fig.1. Schematics of (a) conventional reciprocal coupling and (b) nonreciprocal coupling with nonlocal loss engineering.

## II. THEORETICAL MODELING

The time evolution of a two-port coupled resonator system, as illustrated in Fig. 1(a), is commonly described using Temporal Coupled Mode Theory (TCMT). The governing equations for the complex amplitudes $a_1, a_2$ within resonators 1 and 2 are given by:

$$\begin{cases} \frac{d}{dt}a_1 = (i\omega_1 - \gamma_1 - \gamma_{10})a_1 + i\kappa_{21}a_2 + \sqrt{2\gamma_1}s_{1+} \\ \frac{d}{dt}a_2 = (i\omega_2 - \gamma_2 - \gamma_{20})a_2 + i\kappa_{12}a_1 + \sqrt{2\gamma_2}s_{2+} \end{cases} \quad (1)$$

where $\omega_{i=1,2}$ is the natural resonance angular frequency of resonator $i$. The external loss rate $\gamma_i$ represents energy leakage to the port while the intrinsic loss rate $\gamma_{i0}$, accounting for intrinsic dissipation due to absorption or radiation. The term $\kappa_{ij}$ denotes the coupling coefficient from mode $i$ to mode $j$. In standard linear and time-invariant systems, the principle of reciprocity requires that the coupling coefficients are symmetric ($\kappa_{12} = \kappa_{21} = i\kappa$). $s_{i+}$ is the complex amplitude of the input wave with excitation frequency $\omega$.

The core of our proposed nonlocal scheme, as illustrated in Fig. 1(b), is captured by introducing the modulated nonlocal loss rate to the intrinsic loss rate of resonator 2 as:

$$\gamma'_{20} = \gamma_{20} + \gamma_{nonlocal}(a_1, a_2) \quad (2)$$

where $a_i = a_{i0}e^{i\phi_i}$, $i = 1,2$, with $a_{i0}$ and $\phi_i$ being the amplitude and phase of mode $i$, respectively. This configuration explicitly breaks the spatial symmetry required for reciprocity, moving beyond conventional local nonlinear loss where a resonator's dissipation depends only on its own state. As a proof of concept, we choose a nonlocal modulation form, $\gamma'_{20} = \gamma_{20} + \gamma_k \cdot a_{10}/a_{20}$, which is both analytically tractable and physically illustrative, where $\gamma_k$ is a tunable nonlocal loss coefficient. In this form, further analysis reveals that the system dynamics can be described by an effective Hamiltonian matrix, whose asymmetric off-diagonal terms are the signature of non-reciprocal coupling:

$$\mathcal{H}^{(\text{eff})} = \begin{pmatrix} i\omega_1 - \gamma_1 - \gamma_{10} & i\kappa \\ i\kappa - \gamma_k e^{i\delta\phi} & i\omega_2 - \gamma_2 - \gamma_{20} \end{pmatrix} \quad (3)$$

The phase difference $\delta\phi = \phi_2 - \phi_1$ between the modes can be describe as $\tan^{-1}[(\gamma_{20}\Gamma - \Delta_2^2\gamma_k)/(\Delta_2\Gamma + \Delta_2\gamma_k\gamma_{20})]$ with $\Gamma = \sqrt{\Delta_2^2\kappa^2 + (\gamma_2 + \gamma_{20})^2\kappa^2 - \Delta_2^2\gamma_k^2}$, which depends on system parameters and detuning $\Delta_2 = \omega - \omega_2$. The introduction of this nonlocal term directly breaks the coupling symmetry. To visualize the effect of this engineered asymmetry at the Hamiltonian level, we plot the magnitudes of the forward $|\kappa_{12}|$ and backward $|\kappa_{21}|$ coupling coefficients in Fig. 2 as a function of the modulation coefficient $\gamma_k$ and detuning $\Delta_2$. While we use parameters in the MHz range for illustrative purposes, the proposed mechanism is frequency-agnostic and can be extended to higher optical frequencies. The nonlocal modulation effectively reshapes the system's interaction landscape, transforming the otherwise constant backward coupling into a tunable, curved surface for the forward coupling. In the resonant case ($\Delta_2 = 0$), perfect non-reciprocity is achieved by setting $\gamma_k = \kappa$, which nullifies the forward coupling coefficient ($\kappa_{12} = 0$) while the backward coupling $\kappa_{21}$ remains finite. Furthermore, the forward coupling strength is continuously tunable over the full range from 0 to $\kappa$ by adjusting both the modulation coefficient $\gamma_k$ and the frequency detuning $\Delta_2$.

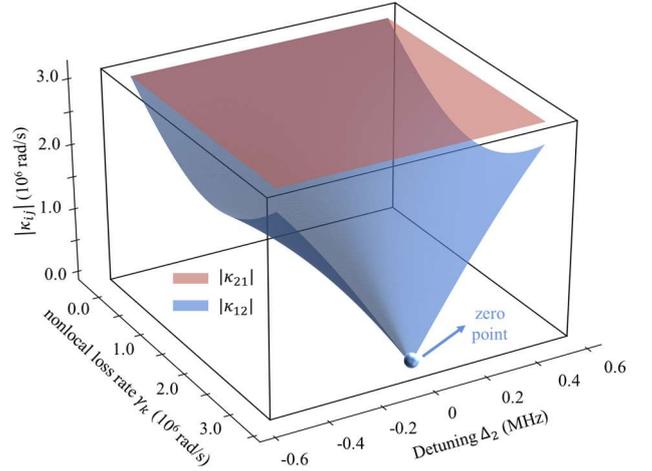

Fig. 2. Visualization of nonreciprocal coupling coefficient (forward $|t_{12}|$, blue surface; backward $|t_{21}|$, red surface) with the simulation parameters $\omega = 1$MHz, $\gamma_{10,20} = 0.5 \times 10^6$ rad/s and $\kappa = \pi \times 10^6$ rad/s.

## III. NURMERICAL RESULTS

To validate our theoretical model and quantify the engineered nonreciprocity, we numerically analyze the system's transmission characteristics by combining and calculating the dynamic equations (1) with time-domain simulations based on the Runge-Kutta method [21].

We begin by exploring the parameter space of our nonlocal system. In Fig. 3, the transmission coefficients $|t_{12}|$ (backward) and $|t_{21}|$ (forward) are plotted as a function of the coupling coefficient $\kappa$ and the nonlocal modulation loss rate $\gamma_k$. As shown in the figure, the simulated data points lie precisely on the theoretical surfaces, demonstrating excellent agreement and validating the accuracy of the effective Hamiltonian described earlier. While both transmission coefficients generally increase with the coupling coefficient $\kappa$, nonlocal modulation $\gamma_k$ introduces pronounced asymmetry. It strongly suppresses forward transmission $|t_{21}|$ while only slightly affecting backward transmission $|t_{12}|$.

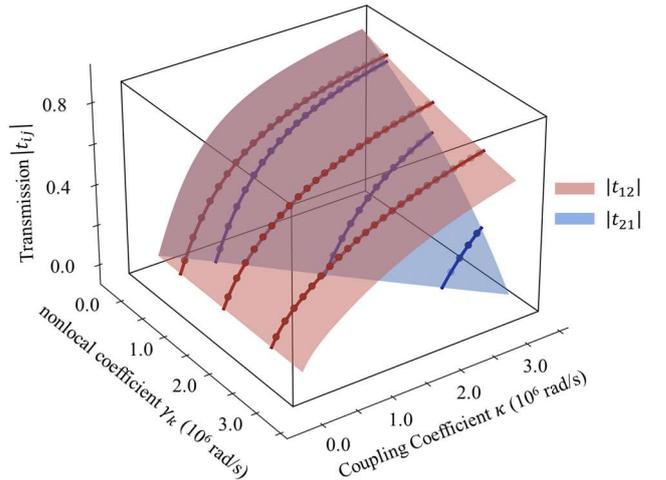

Fig. 3. Comparison between the theoretical model and numerical simulations. The transparent surfaces show the theoretical transmission coefficients for forward ($|t_{12}|$, red) and backward ($|t_{21}|$, blue) propagation. The overlaid solid lines are cross-sections illustrating the transmission dependence on the coupling coefficient $\kappa$ at fixed values of the modulation coefficient $\gamma_k$. Discrete points represent numerical simulation results. Simulations are performed at resonance ($\Delta_2 = 0$) with $\gamma_{10} = \gamma_{20} = 0.5 \times 10^6$ rad/s and $\gamma_i = \sqrt{\gamma_{i0}^2 + \kappa^2}$ (for minimal reflection).

As $\gamma_k$ increases, the disparity between forward and backward transmission grows, visualized by expanding contrast between the two surfaces at identical parameter points—signifying the onset and enhancement of nonreciprocity. This nonreciprocal behavior leads to a striking feature: when $\gamma_k = \kappa$, the forward transmission is completely suppressed, achieving perfect isolation ($|t_{21}| = 0$), while the backward transmission $|t_{12}|$ remains finite and can be tuned via $\kappa$ and $\gamma_k$.

Next, we turn to the frequency domain to assess the robustness of our system against detuning. Fig. 4 shows that a consistently high transmission contrast between backward and forward directions is maintained over a 40% 3dB relative bandwidth, with a center frequency set at 1 MHz for demonstration; in principle, this approach can be extended to arbitrarily higher frequencies. This persistence not only attests to the system's broadband capability but also underscores its stability against practical frequency fluctuations.

Moving beyond steady-state analysis, we now probe the time-domain dynamics to further capture the essence of nonreciprocal performance. As shown in Fig. 5(a), when a signal is injected from port 1, our nonlocal system exhibits perfect forward isolation, with a response that is indistinguishable from that of an ideal nonreciprocal device. Conversely, as shown in Fig. 5(b), when the system is excited from port 2, a slight increase in reflection is observed at the input port; nevertheless, the output waveform at port 1 remains in excellent agreement with the ideal model. These results demonstrate how nonlocal loss engineering empowers a reciprocal system to faithfully emulate ideal nonreciprocal coupling process in real time.

Finally, we investigate the robustness of our design against input intensity variations. Fig. 6 highlights the remarkable power independence of our nonlocal approach: isolation remains above 20 dB across the tested range, in sharp contrast to local nonlinear loss models whose performance is highly sensitive to signal strength. Such robustness enables our scheme to maintain high isolation even under fluctuating input conditions. This power-independent stability is also clearly supported by our theoretical model, which yields the bidirectional transmission efficiencies $\eta_{12}$ (from port 2 to 1) and $\eta_{21}$ (from port 1 to 2) as:

$$\eta_{21} = \frac{4\gamma_1\gamma_2|i\kappa - \gamma_k e^{i\delta\phi}|^2}{|\det(\mathcal{H}^{(\text{eff})} - \omega I)|^2},$$
$$\eta_{12} = \frac{4\gamma_1\gamma_2\kappa^2}{|\det(\mathcal{H}^{(\text{eff})} - \omega I)|^2} \quad (4)$$

where $I$ is the identity matrix, and the determinant $\det(\mathcal{H}^{(\text{eff})} - \omega I) = (i\Delta_1 + \gamma_1 + \gamma_{10})(i\Delta_2 + \gamma_2 + \gamma_{20}) + \kappa^2 + i\kappa\gamma_k e^{i\delta\phi}$. Equation (4) shows that, unlike conventional local nonlinear loss schemes—which intrinsically link transmission to input intensity—our nonlocal loss engineering strategy effectively decouples transmission properties from input power.

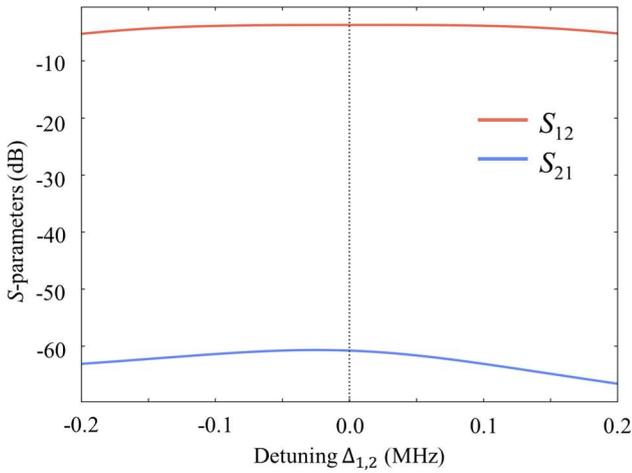

Fig. 4. *S*-parameters against frequency detuning. The simulation is performed at a point of perfect isolation where the modulation coefficient and coupling rate are matched ($\gamma_k = \kappa = 0.3 \times 10^6$ rad/s). Other parameters are the same as mentioned previously.

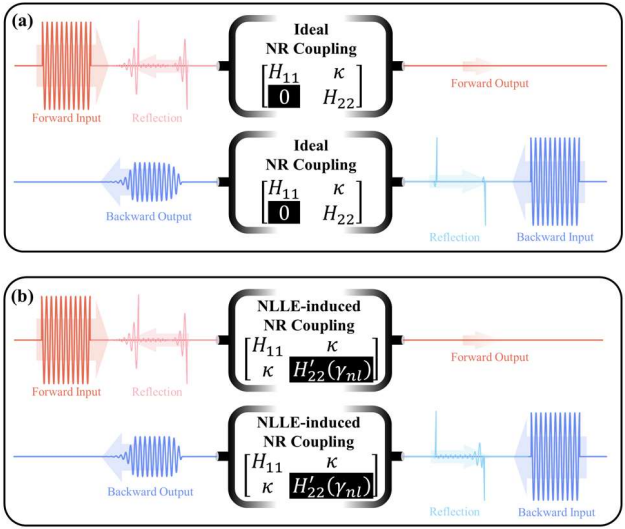

Fig. 5. Time-domain comparison of ideal and NLLE-induced nonreciprocal systems. Numerically calculated response of (a) an ideal nonreciprocal (NR) system and (b) our nonlocal loss-engineered (NLLE-induced) system to forward (red) and backward (blue) inputs. $\gamma_{nl}$ represents the nonlocal loss. In both panels, the transmitted wave is shown in the same color as the input, while the reflected wave is depicted in a lighter shade. For the forward input, the red transmitted wave exhibits zero amplitude in both systems, signifying the achievement of perfect non-reciprocity.

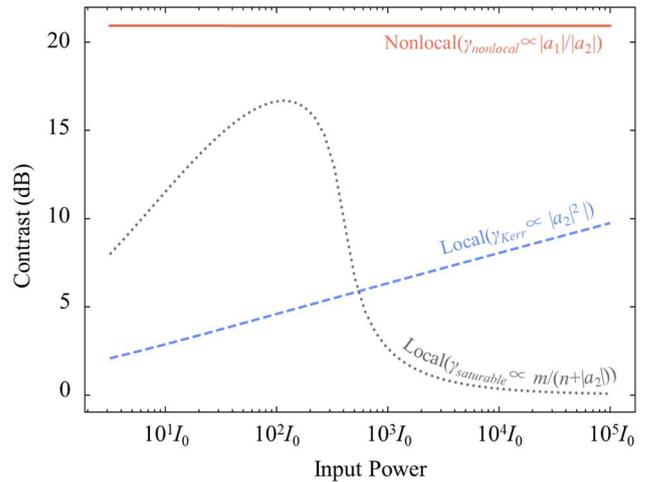

Fig. 6. Stability against input power variations. Comparison of the isolation performance for our nonlocal model against two common local models as a function of input power. The solid red line represents our nonlocal model ($\gamma_{nonlc} \propto |a_{10}|/|a_{20}|$), which maintains a constant high isolation across all power levels. The dashed blue line represents a local Kerr-like nonlinear loss model ($\gamma_{Kerr} \propto |a_{20}|^2$), and the dotted gray line represents a local model with saturable nonlinear loss ($\gamma_{saturable} \propto |a_{20}|^2$). Simulations is performed on resonance ($\Delta = 0$) with $\kappa = 3 \times 10^6$ rad/s and $\gamma_{10} = \gamma_{20} = 0.5 \times 10^6$ rad/s.

## IV. Conclusion

In summary, we have proposed and numerically validated a mechanism for nonreciprocal coupling based on nonlocal loss engineering. By dynamically modulating the loss in one optical mode according to the state of another, we achieve an asymmetric coupling coefficient and realize broadband, high-contrast nonreciprocity. Simulations further confirm that this nonreciprocal performance is stable across all input power levels, indicating that nonlocality is key to overcoming the intrinsic power-dependent limitations of conventional nonlinear schemes. Although implementation in the infrared and visible spectra requires ultrafast and precise loss control—potentially utilizing graphene transistors, flash ionization in plasmas, or giant optical nonlinearities in epsilon-near-zero materials [22]—this concept can be readily demonstrated using microwave circuits. Our results establish a conceptual foundation for a new class of active, tunable, magnetic-free nonreciprocal devices. These can enable novel components—such as free-space photonic isolators and circulators—and offer a platform for exploring unexploited phenomena, including long-range nonreciprocal light-matter interactions. The tunable and integrable nature of this concept also makes it promising for simulating large-scale neural networks and investigating topological effects such as high-dimensional band braiding and high-order non-Hermitian skin effect.